\documentclass[aps,prd,twocolumn,superscriptaddress,preprintnumbers]{revtex4}

\usepackage{epsf,epsfig,amsmath,amssymb,amsfonts}
\usepackage{color}
\usepackage{slashed}

\newsavebox{\ns}
\newsavebox{\dbrane}

\def\be{\begin{equation}}
\def\ee{\end{equation}}
\def\bea{\begin{eqnarray}}
\def\eea{\end{eqnarray}}

\def\Dslash{\,\,{\raise.15ex\hbox{/}\mkern-12mu D}}
\def\Dbarslash{\,\,{\raise.15ex\hbox{/}\mkern-12mu {\bar D}}}
\def\delslash{\,\,{\raise.15ex\hbox{/}\mkern-9mu \partial}}
\def\delbarslash{\,\,{\raise.15ex\hbox{/}\mkern-9mu {\bar\partial}}}
\def\pslash{\,\,{\raise.15ex\hbox{/}\mkern-9mu p}}
\def\calDslash{\,\,{\raise.15ex\hbox{/}\mkern-12mu {\cal D}}}

\newcommand\diff{\mbox{d}}

\newcommand{\nn}{\nonumber \\}

\newcommand{\dd}{\diff}

\newcommand{\adss}{AdS$_5\times$S$^5$ }

\allowdisplaybreaks

\begin{document}

\title{Yang-Baxter $\sigma$-models, conformal twists, and noncommutative Yang-Mills theory}

\preprint{APCTP Pre2017 - 003}         

\preprint{DMUS-MP-17/03}   

\preprint{KUNS-2663}

\preprint{IPM/P-2017/006}
\vskip 1 cm

 \author{T. Araujo}
 \affiliation{Asia Pacific Center for Theoretical Physics, Postech, Pohang 37673, Korea}
 \author{I. Bakhmatov}
 \affiliation{Asia Pacific Center for Theoretical Physics, Postech, Pohang 37673, Korea}
 \affiliation{Institute of Physics, Kazan Federal
University, Kremlevskaya 16a, 420111, Kazan, Russia}
 \author{E. \'O Colg\'ain}
 \affiliation{Asia Pacific Center for Theoretical Physics, Postech, Pohang 37673, Korea}
 \affiliation{Department of Mathematics, University of Surrey, Guildford GU2 7XH, United Kingdom}
\author{J. Sakamoto}
 \affiliation{Department of Physics, Kyoto University,  Kitashirakawa, Kyoto 606-8502, Japan}
\author{M. M. Sheikh-Jabbari}
\affiliation{School of Physics, Institute for Research in Fundamental Sciences (IPM), P.O.Box 19395-5531, Tehran, Iran}
\author{K. Yoshida}
 \affiliation{Department of Physics, Kyoto University,  Kitashirakawa, Kyoto 606-8502, Japan}

\begin{abstract}
\noindent
The Yang-Baxter $\sigma$-model is a systematic way to 
generate integrable deformations of AdS$_5\times$S$^5$. We recast the deformations as seen by open strings, where the metric is undeformed AdS$_5\times$S$^5$ with constant string coupling, and all information about the deformation is encoded in the noncommutative (NC) parameter $\Theta$. We identify the deformations of AdS$_5$ as twists of the conformal algebra, thus explaining the noncommutativity. We show that the unimodularity condition on $r$-matrices for supergravity solutions translates into $\Theta$ being divergence-free. Integrability of the $\sigma$-model for unimodular $r$-matrices implies the existence and planar integrability of the dual NC gauge theory.

\end{abstract}

\maketitle

\setcounter{equation}{0}

\section{Introduction} \label{Introduction}
Integrable models have been key to enriching our knowledge of condensed matter systems, field theory, and string theory. Within string theory, considerable attention has focused on integrable structures underlying the AdS (Anti de Sitter)/CFT (Conformal Field Theory) correspondence \cite{Maldacena:1997re}. The most studied example is a duality between superstrings on AdS$_5 \times$S$^5$ and $\mathcal{N} = 4$ super Yang-Mills (sYM). Remarkably, the two-dimensional string world sheet $\sigma$-model on \adss is classically integrable \cite{Bena:2003wd};  it has an infinite set of conserved charges.  

There is immense interest in identifying integrable structures beyond the maximally symmetric setting of AdS$_5\times$S$^5$, or equivalently sYM on $\mathbb{R}^{1,3}$. It is curious that the earliest integrability preserving deformation of AdS$_5\times$S$^5$ \cite{Hashimoto:1999ut, Maldacena:1999mh, Alishahiha:1999ci} was inspired by noncommutative (NC) spacetimes, which are ubiquitous in string theory  \cite{Ardalan:1998ce, Seiberg:1999vs} (see \cite{Szabo:2001kg} for a review). In hindsight, we understand these deformations as T-duality shift T-duality (TsT) transformations {in the string and gravity side}  \cite{Lunin:2005jy, Frolov:2005dj, Frolov:2005ty}. 

Recently, Yang-Baxter(YB) deformations of the $\sigma$-model \cite{Klimcik:2002zj, Klimcik:2008eq, Delduc:2013fga,Matsumoto:2015jja} were generalized to the AdS$_5\times$S$^5$ superstring 
\cite{Delduc:2013qra,Kawaguchi:2014qwa}. We now understand  TsT transformations as part of 
a larger class of YB deformations of the $\sigma$-model \cite{Matsumoto:2014nra, Matsumoto:2015uja, Matsumoto:2014gwa, Matsumoto:2014ubv,Kawaguchi:2014fca,vanTongeren:2015soa,vanTongeren:2015uha,Kyono:2016jqy,
Hoare:2016hwh,Orlando:2016qqu,Borsato:2016pas,Osten:2016dvf,vanTongeren:2016eeb,Hoare:2016wca},  
which are defined by $r$-matrices satisfying the classical Yang-Baxter equation (cYBE). A further unimodularity condition ensures the YB deformation has a valid string theory (supergravity) description \cite{Borsato:2016ose}. 
It has been conjectured \cite{Hoare:2016wsk} (see also \cite{Borsato:2016pas}) that homogeneous YB deformations 
\cite{Matsumoto:2015jja,Kawaguchi:2014qwa} may all be realized through non-Abelian duality transformations \cite{Fridling:1983ha, Fradkin:1984ai, delaOssa:1992vci, Sfetsos:2010uq, Lozano:2011kb}. 

In this article, we retrace TsT transformations to NC deformations of quantum field theories (QFTs). We encounter a number of surprises. First, irrespective of the YB deformation, for $r$-matrix solutions to the homogeneous cYBE, there is a universal description in open string parameters. Concretely, we show that the open string metric \cite{Seiberg:1999vs} is always the original \textit{undeformed} \adss metric with constant open string coupling, and all information about the YB deformation is encoded in a NC parameter $\Theta$. This in particular implies that all YB string theory $\sigma$-models of \adss have a NC gauge theory dual on $\mathbb{R}^{1,3}$ where integrability of the $\sigma$-model has direct bearing on planar integrability. 

For our second result, sharpening an earlier conjecture \cite{vanTongeren:2015uha}, we confirm that YB deformations of  AdS$_5$ are simply Drinfeld twists of the conformal algebra. To better understand this fact, we recall that in NC spacetimes the coordinate operators $\hat{x}^{\mu}$ satisfy the commutation relation,  
\be\label{NC_spacetime}
\left[ \hat{x}^{\mu}, \hat{x}^{\nu} \right] = i \Theta^{\mu \nu} \qquad (\mu,\nu=0,\ldots,3), 
\ee
where $\Theta^{\mu \nu}$ is in general an $x$-dependent antisymmetric matrix. For twists of  Poincar\'e algebra, the $x$-dependence of $\Theta$ is fixed to be constant, linear or quadratic
\cite{Chaichian:2004za, Chaichian:2004yh, Lukierski:2005fc}. As we will argue, however, for twists in the conformal algebra we can also have cubic and quartic dependence. In fact, the homogeneous YB deformations studied to date \cite{Matsumoto:2014nra, Matsumoto:2015uja, Matsumoto:2014gwa, Matsumoto:2014ubv,Kawaguchi:2014fca,vanTongeren:2015soa,vanTongeren:2015uha,Kyono:2016jqy,
Hoare:2016hwh,Orlando:2016qqu,Borsato:2016ose,Osten:2016dvf,vanTongeren:2016eeb,Hoare:2016wca}
provide predictions for NC parameters that arise from twists of the \textit{full} conformal algebra. We establish by exhaustion that the NC parameters and $r$-matrices are directly related \cite{inprogress}, 
\be\label{Theta-r}
\Theta^{MN} = -2 \, \eta \, r^{MN} \qquad (M,N=0,\ldots,3,z), 
\ee
where $\eta$ is the deformation parameter, $z$ is the radial direction of AdS$_5$, and $r^{MN}$ is the $r$-matrix expressed as differential operators on AdS$_5$.

Finally, non-unimodular YB deformations lead to geometries that solve generalized supergravity equations, specified through a modification given by a Killing vector field $I$ \cite{Arutyunov:2015mqj,Wulff:2016tju}; setting $I = 0$, we recover usual supergravity. We show $\Theta$ and $I$ are related through the equation, 
\be\label{unimodular}
\nabla_{M} \Theta^{MN} = I^{N}, 
\ee  
evaluated with an open string metric. 
This remarkable result, which marries open and closed string descriptions, is a requirement of the $\Lambda$-symmetry \cite{Witten:1995im,SheikhJabbari:1999ba} of the string $\sigma$-model. Under $\Lambda$-symmetry the NSNS two-form $B$-field is transformed by $\dd \Lambda$, which  in the presence of D-branes (open strings) must be supplemented by a shift of the gauge field on the brane by a one-form $\Lambda$.  This novel observation provides the first explanation of the unimodularity condition \cite{Borsato:2016ose} from a symmetry principle. Observe, for supergravity solutions, $\Theta^{MN}$ is divergence-free.

\section{closed string picture}
In an effort to make this article self-contained, we review the essentials of the YB $\sigma$-model, following the presentation of Ref.\,\cite{Kyono:2016jqy}. Here,  
we restrict ourselves to deformations of AdS$_5$ by considering the coset space $SO(4,2)/SO(4,1)$ and the homogeneous cYBE. Furthermore, to avoid unnecessary technicalities, we suppress the RR sector, which does not affect any of our results. 
The corresponding YB  $\sigma$-model action is \cite{Matsumoto:2015jja,Kawaguchi:2014qwa}
\be
\mathcal{L} = \textrm{Tr} \left[ A\,P^{(2)} \circ \frac{1}{1 - 2 \eta R_g \circ P^{(2)}} A\right], 
\ee
with a deformation parameter $\eta$ and $R_{g} (X) \equiv g^{-1} R(g X g^{-1}) g$\,.
Here $A=-g^{-1}\dd g$, $g\in SO(4,2)$, is a left-invariant current, while
$P^{(2)}$ is a projector onto the coset space $\mathfrak{so}(4,2)/\mathfrak{so}(4,1)$, spanned by the generators ${\bf P}_{m}\,(m=0,\ldots,4)$, which satisfy $\textrm{Tr}[{\bf P}_{m}{\bf P}_{n}]=\eta_{mn} =\textrm{diag}(-++++)$.  Details, such as matrix representations, are given in \cite{Kyono:2016jqy}. $P^{(2)}$ may be expressed as
\be
P^{(2)}(X)=\eta^{mn}{\rm Tr}[X\,{\bf P}_{m}]\,{\bf P}_{n}\,,\quad X\in \frak{so}(4,2)\,.
\ee 

Above, $R$ is an antisymmetric operator satisfying the homogeneous cYBE 
\be
\label{cYBE}
[ R (X), R (Y)] - R([R (X), Y]+ [X, R (Y)])=0,   
\ee
with $X, Y \in \frak{so}(4,2)$. 
In turn, the operator $R$ can be written in terms of an $r$-matrix as 
\be
R(X)  =\textrm{Tr}_2[r(1\otimes X)] =\sum_{i,j}r^{ij} b_i \textrm{Tr} [ b_j X],
\ee
where $r\in \frak{so}(4,2)\otimes \frak{so}(4,2)$ is 
\be
r =\frac12 \sum_{i,j}  r^{ij} b_i\wedge b_j, \quad {\rm with}\quad b_i\in \frak{so}(4, 2).
\ee
The $r$-matrix is called Abelian if $[b_i,b_j]=0$ and unimodular if 
it satisfies the following condition \cite{Borsato:2016ose}: 
\be\label{unimodular_r}
r^{ij}[b_i,b_j]=0.
\ee
Note $i, j$ range over the generators of $\frak{so}(4,2)$, but expressed as differential operators on AdS$_5$, one finds $r^{MN}$. 

To determine the YB deformed geometry, we adopt the following parametrization for $g \in SO(4, 2)$:
\be
g = \exp [ x^{\mu} P_{\mu} ] \exp [ (\log z) D], 
\ee
where $P_{\mu}\,(\mu=0,...,3)$, $D$, respectively denote translation and dilatation generators and are related to ${\bf P}_{m}$ \cite{Kyono:2016jqy}. In terms of these coordinates, we define 
\be
r = \frac{1}{2} r^{MN} \partial_{M} \wedge \partial_{N}, \quad \partial_{M} \in \{ \partial_{\mu}, \partial_{z} \}. 
\ee 
Then, the YB deformed metric $g_{MN}~(M,N=0,\ldots,4)$, NSNS two-form $B_{MN}$, and dilaton $\Phi$  (in string frame) can be expressed as \cite{Kyono:2016jqy} 
\bea
\label{YB_deformation}
g_{MN} &=& e^{m}_{M} e^{n}_{ N} k_{(mn)},
\quad B_{MN} = e^{m}_{M} e^{n}_{N} k_{[nm]}, \\
\label{dilaton} {\rm e}^\Phi &=& g_s(\textrm{det}_{5}\,k)^{-1/2}\,,\ \ k_{mn}  = k_{(mn)} + k_{[mn]},
\eea
where $e^{m}_{M}$ is the AdS$_5$ vielbein, and we have defined
\bea
\label{k} k_{m}{}^{n}&\equiv& ( \delta^{m}{}_{n} - 2 \eta \lambda^{m}{}_{n} )^{-1}, \\
\label{lambda} \lambda_{m}{}^{n} &\equiv&
 \eta^{nl}{\rm Tr}[{\bf P}_{l}R_g({\bf P}_{m})]. 
\eea

It is useful to exemplify the deformation for the simplest case of the Abelian $r$-matrix \cite{Matsumoto:2014gwa},
\be
r= \frac{1}{2}P_2\wedge P_3,
\ee
corresponding to the closed string background \cite{Hashimoto:1999ut,Maldacena:1999mh}, 
\bea
\label{MR-soln}
\dd s^2&=&\frac{1}{z^2}[-\dd x_0^2+\dd x_1^2+h(z)(\dd x_2^2+\dd x_3^2)+\dd z^2]
\nonumber\\
B_{23}&=&\eta h(z)/z^4, \quad {\rm e}^{2\Phi}=g_s^2 h(z),
\eea
where $h^{-1}=1+\eta^2 z^4$. The above together with {S$^5$ and} the RR-fields constitute a supergravity solution, 
which is obtained simply via TsT from \adss
\cite{Hashimoto:1999ut,Maldacena:1999mh}.

In passing, we comment that while we focus on AdS$_5$, following \cite{Matsumoto:2014nra}, similar arguments apply equally to S$^5$. In particular,  the case of $\beta$ \cite{Lunin:2005jy} or $\gamma$-deformations \cite{Frolov:2005dj} is related to Abelian twists of $SO(6)$, and via AdS/CFT, to marginal deformations of $\mathcal{N}=4$ sYM \cite{Leigh:1995ep}.

\section{Open string picture} 
Given closed string parameters ($g_{MN}$, $ B_{MN}$, $g_s$),   
the open string metric $G_{MN}$, NC parameter $\Theta^{MN}$ and coupling $G_s$ are defined as \cite{Seiberg:1999vs} 
\bea
G_{MN} &=& \left( g- Bg^{-1}B\right)_{MN},\label{Open-metric} \\
\Theta^{MN} &=& - \left( ({g + B})^{-1} B ({g - B})^{-1} \right)^{MN},\label{Theta-def} \\
 G_s &=& g_s {\rm e}^{\Phi} \left( \frac{\det (g+B)}{ \det g} \right)^{\frac{1}{2}} \label{Gs}.
\eea

For YB deformations of AdS$_5$ \eqref{YB_deformation}, we find 
\be
G^{MN} + \Theta^{MN} = e^{M}_{m} e^{N}_{n} \left( \eta^{mn} + 2 \eta \, 
\lambda^{mn} \right), 
\ee
where $e^{M}_{m}$ denotes the inverse vielbein. As $\lambda^{mn}$ is antisymmetric, it is easy to separate the components, getting 
\be
\label{G_theta}
G^{MN} = e^{M}_{m} e^{N}_{n} \eta^{mn}, \quad \Theta^{MN} =  2 \eta \,e^{M}_{m} e^{N}_{n} \, \lambda^{mn}. 
\ee
Inverting $G^{MN}$,  it is clear that the open string metric is precisely the original AdS$_5$ metric. Moreover, inserting \eqref{YB_deformation} and \eqref{dilaton} into \eqref{Gs}, we get $G_s=g_s = {\rm const}$. That is, all the information about the YB deformation, as viewed by open strings, is sitting in $\Theta^{MN}$, while the geometry is undeformed AdS$_5$
\footnote{It is easy to see via ansatz, e.\,g.\,\cite{Colgain:2016gdj}, for general TsT transformations that the open string metric and coupling are just the pre-TsT closed string metric and coupling.}. 

For the example \eqref{MR-soln}, the open string parameters are 
\bea 
\label{MR-case-Open} 
\dd s^2_{\rm open} &=& \frac{1}{z^2}({-\dd x_0^2 + \dd x_1^2 + \dd x_2^2 + \dd x_3^2 + \dd z^2 }), 
\nn
\Theta^{23}&=& -\eta, \qquad G_s=g_s.
\eea
While the closed string metric \eqref{MR-soln} has a severely deformed causal and boundary structure \cite{Hashimoto:1999ut, Maldacena:1999mh, Alishahiha:1999ci}, the spacetime as seen by the open strings is the usual \adss with $\mathbb{R}^{1,3}$ boundary, indicating that the dual gauge theory description is a $\Theta$-deformed sYM.

\section{Conformal twists and NC gauge theory }
One can formulate the QFT on the NC spacetime specified by $\Theta$ \eqref{NC_spacetime}. Let us start with the constant $\Theta$ case, relevant to the example (\ref{MR-soln}). The NCQFT may be obtained by replacing the usual product of functions, or fields in QFT, with the Moyal star product, $f(x) g(x) \rightarrow  (f \star g)(x)$, such that 
\be
\label{moyal_star}
(f \star g) (x) = f(x) {\rm e}^{\frac{i}{2} \Theta^{\mu \nu} \overset{\leftarrow}{\partial_{\mu}} \overset{\rightarrow}{{\partial}_{\nu}} } g(x).  
\ee
The Moyal bracket of two functions is defined to be 
\be
\label{moyal_bracket}
[f, g]_{\star} := f \star g - g \star f = i \Theta^{\mu \nu} \partial_{\mu} f \partial_{\nu} g + \mathcal{O} ( \partial^3 f, \partial^3 g). 
\ee
It is worth noting that $f(x) = x^{\mu}, g(x) = x^{\nu}$ reproduces the commutator (\ref{NC_spacetime}). It has been shown that the introduction of the Moyal $\star$-product is equivalent to using the coproducts with a Drinfeld twist element \cite{Chaichian:2004za}, 
\be
\label{twist}
\mathcal{F} = {\rm e}^{-2i\eta r} = {\rm e}^{ \frac{i}{2} \Theta^{\mu \nu} P_{\mu} \wedge P_{\nu} }.  
\ee 
This is a special case of an Abelian Poincar\'e twist, and the $r$-matrix satisfies the cYBE \cite{Matsumoto:2014gwa}. Abelian twists have the remarkable property that they do not affect the Poincar\'e algebra $\mathcal{P}$ \cite{Chaichian:2004za}, but instead deform the coproduct of $\mathcal{U} (P)$ \cite{Drinfeld}, where $\mathcal{U} (P)$ is the universal enveloping algebra of the Poincar\'e algebra.

In \eqref{twist}, we have considered the simplest twist, with constant $\Theta$.  However, for other solutions to the cYBE, the NC parameter need not be a constant. Indeed, including Lorentz generators $M_{\mu\nu}$, the cYBE has solutions $r \sim P \wedge M$ and $r \sim M \wedge M$, which, respectively, lead to
linear and quadratic $\Theta$ \cite{Lukierski:2005fc}. 
For example, for $ r = \frac{1}{2} M_{01} \wedge M_{23}$, modulo a convention dependent sign in the twist (\ref{twist}), the NC parameter has components \cite{Lukierski:2005fc} 
\be\begin{split}
\Theta^{02} =& - 2 \sinh \frac{\eta}{2} \cdot x^1 x^3, \quad
 \Theta^{03} =  2 \sinh \frac{\eta}{2} \cdot x^1 x^2, \\
\Theta^{12} =& - 2 \sinh \frac{\eta}{2} \cdot x^0 x^3, \quad
\Theta^{13} =  2 \sinh \frac{\eta}{2} \cdot x^0 x^2. 
\end{split}
\ee
We recover the same result (at leading order) from the YB prescription (\ref{G_theta}). 

This example shows that the open string parameter $\Theta$ knows about the Moyal bracket, which may be derived from twists of the Poincar\'e algebra. One can repeat the YB analysis for \emph{all} $r$-matrices of the \emph{conformal algebra} and show that (\ref{Theta-r}) holds once the $r$-matrix is expressed in terms of differential operators \cite{inprogress}. Note, (\ref{Theta-r}) generalizes existing results \cite{vanTongeren:2015uha, vanTongeren:2016eeb} from the Poincar\'e to conformal algebra.
 
In support of our claim, we present two examples
\be\begin{split}
\label{r1,2} r_1 =& \frac{1}{2}D \wedge K_1, \\
 r_2 =& \frac{1}{2}  (P_0 - P_3) \wedge (D + M_{03}), 
\end{split}
\ee
which involve scale $D$ and special conformal symmetries $K_{\mu}$. Note, the first is non-unimodular and the second appears in the classification of unimodular $r$-matrices \cite{Borsato:2016ose}. 
The NC parameter in each case can easily be calculated from (\ref{lambda}) and (\ref{G_theta}). For $r_1$, we find
\bea\label{Theta-r1}
\Theta^{1 {\mu}} &=& \eta x^{{\mu}} ( x_{\nu} x^{\nu}  + z^2), \quad \Theta^{1z} = \eta z ( x_{\nu} x^{\nu}  + z^2), 
\eea 
where $\mu \neq 1$, while for $r_2$, we get 
\bea\label{Theta-r2}
\Theta^{-+} = -4 \eta x^+, \quad \Theta^{-i} = -2 \eta x^i, \quad \Theta^{- z} = -2 \eta z, 
\eea
where $i = 1, 2$ and we have employed $x^{\pm} = x^0 \pm x^3$. One recovers the same results from conformal twists of the dual CFT \cite{inprogress}. We interpret this mathematical agreement as evidence in support of our claim that YB deformations based on unimodular $r$-matrices are dual to NC deformations of $\mathcal{N}=4$ sYM. We establish this through an almost exhausting set of examples in our upcoming work \cite{inprogress}. 

Some comments and remarks are in order:
\begin{itemize}
\item[1)] In both cases one can confirm that Eq.\,(\ref{Theta-r}) holds. 
\item[2)] One generically encounters cubic and quartic terms from conformal twists. 
\item[3)] Not only are there nonzero $\Theta^{z\mu}$ components, they also have nontrivial $z$-dependence. Nonetheless, it can be shown in general that $\Theta^{z\mu}$ components vanish at the AdS boundary at $z=0$, where the dual field theory resides. Viewing Eq.\,\eqref{unimodular} as a first order equation for $\Theta^{MN}$,  the $z$-components and dependence can be recovered from the $\Theta^{\mu\nu}$; no information is lost in the dual field theory side.
\item [4)] For YB deformations corresponding to unimodular $r$-matrices, there is a well-defined string theory picture. Following the usual reasoning of AdS/CFT, wherever the decoupling limit exists, closed string theory on these deformed AdS$_5$ backgrounds is expected to be dual to NC deformations of sYM with noncommutativity $\Theta^{\mu \nu} = -2\eta r^{\mu\nu}$. Particular examples are discussed in  \cite{Hashimoto:1999ut, Maldacena:1999mh, Alishahiha:1999ci}. However, we note that the existence of a decoupling limit, where the open string theory is reduced to its low energy limit of NC sYM, is not trivial \cite{Alishahiha:1999ci} (see also \cite{vanTongeren:2015uha, vanTongeren:2016eeb} for related discussion). For the cases with $\Theta^{\mu\nu}\Theta_{\mu\nu}<0$, so-called ``electric'' noncommutativity, it has been argued that the open string theory does not reduce to NC sYM. In these cases we are dealing with the noncritical NC open string theory (NCOS) \cite{NCOS1,NCOS2,NCOS3,NCOS4} which is related to  NC sYM at strong coupling.
\end{itemize}

\section{Unimodularity and $\Lambda$-symmetry} 
Our statements about the universal open string description are true, irrespective of unimodularity. Here, we address the origin of unimodularity in terms of string theory and its symmetry. 

The key to our explanation is $\Lambda$-symmetry  \cite{Witten:1995im,SheikhJabbari:1999ba}. It is known that closed string theory (supergravity) is invariant under $B\to B+\dd \Lambda$, where $B$ is the NSNS two-form and $\Lambda$ is an arbitrary one-form. Upon introduction of open strings with Dirichlet boundary conditions, this symmetry survives, since $B$ appears in the brane DBI action only through the combination $B+F$, where $F = \dd A$ is the field strength of the brane gauge field $A$ \cite{Leigh:1989jq}, and one can compensate by shifting $A\to A-\Lambda$. Therefore, the action of the system, which is the sum of the supergravity and DBI actions, maintains the $\Lambda$-symmetry. 

Open string parameters \eqref{Open-metric}, \eqref{Theta-def}, and \eqref{Gs}, however, are defined in a particular $\Lambda$-gauge, where the expectation (or background) value of $F$ is set to zero. So, the expression for $\Theta^{MN}$ \eqref{Theta-def} is not necessarily $\Lambda$-invariant  \cite{Seiberg:1999vs,SheikhJabbari:1999ba}. In fact, recalling that when $F$ is set to zero \cite{Seiberg:1999vs},
$$
\frac{1}{G_s}\sqrt{\det{G}}=\frac{{\rm e}^{\Phi}}{g_s}\sqrt{\det{(g+B)}},
$$
one can readily see that the variation of the DBI action with respect to $\Lambda$-symmetry is $\nabla_M \Theta^{MN}$, where the divergence is computed with respect to open string metric $G_{MN}$. So, invariance of the full action for the unimodular cases where the supergravity part is $\Lambda$-invariant on its own leads to $\nabla_M\Theta^{MN}=0$. See also \cite{Szabo:2006wx} for related arguments.

For the non-unimodular cases,  where we encounter generalized supergravity equations with Killing vector $I$, one can show that these equations are $\Lambda$-symmetric. However, the presence of the isometry direction $I$ would modify the DBI action by an $I^MA_M$ term, which is not $\Lambda$-invariant \cite{inprogress}. Therefore, to restore $\Lambda$-symmetry, the NC parameter should satisfy  \eqref{unimodular}. As an example consider $r_1$ in \eqref{r1,2}, which  is known to be non-unimodular with $I=K_1$. One can then  explicitly check that $\Theta$ given in \eqref{Theta-r1} satisfies \eqref{unimodular}.

\section{Outlook}
Our observations and results have broad implications. It is imperative to revisit Poincar\'e twists \cite{Chaichian:2004za, Chaichian:2004yh, Lukierski:2005fc} and extend them to conformal twists \cite{inprogress}, thus testing our claim that
the conformal twists can be described as YB deformations. While we considered only bosonic deformations of AdS$_5$, one can easily repeat for different coset spaces, in different dimensions, or extend the analysis to the fermionic sector of the \adss $\sigma$-model, where one will encounter fermionic T-duality \cite{Berkovits:2008ic, Beisert:2008iq}, or potentially a non-Abelian generalization of it.

We recall that the homogeneous YB deformations may be described as non-Abelian T-duality \cite{Hoare:2016wsk} . In principle, a careful treatment of the $\Theta$ parameter for non-Abelian T-duals supported by RR flux \cite{Sfetsos:2010uq, Lozano:2011kb} may elucidate the dual theory \footnote{Recent attempts to identify CFTs dual to non-Abelian T-duals have focussed on linear quivers \cite{Lozano:2016kum, Lozano:2016wrs}.}.  It is interesting that the open string, via $\Lambda$-symmetry, knows about generalized supergravity through $I$. Since the latter is reproducible from the Double Field Theory description, it may be interesting to push this connection by following \cite{Sakatani:2016fvh, Baguet:2016prz,Sakamoto:2017wor}.

The \adss 
 YB $\sigma$-model integrability has implications for the dual  gauge theory and the dual open strings. The fact that open strings reside in an undeformed \adss geometry prompts the proposal of  integrability of the corresponding open string $\sigma$-model. The effects of the deformation should then appear in $\Theta$ which is expected to affect only open string end point dynamics (which end on the \adss boundary). This open string integrability dovetails with the fact that {some} of the deformed backgrounds can be obtained through TsT transformations and that T-duality is a symmetry of the world sheet theory. Establishing this open string integrability proposal, however, requires a thorough analysis of the boundary conditions.
 
Integrability of the \adss $\sigma$-model is intimately connected with the planar integrability of the corresponding dual ${\cal N}=4$ sYM. With the same token, one would expect the associated NC sYM to be planar integrable. Some preliminary analysis and results for a special case have already appeared \cite{Beisert:2005if}. This is a highly nontrivial statement and extends the important sYM integrability to a big list of NC gauge theories. In the same line, one would expect that Drinfeld twists and Drinfeld doubles of the original Yangian, which underlies the integrability of sYM, to be at work for the NC cases. 

It is known that the constant magnetic NC sYM at strong coupling flows to the NCOS \cite{NCOS1,NCOS2,NCOS3,NCOS4}. It is interesting to check if the same feature extends to more general $x$-dependent twist elements. Recalling the S-duality of 
type IIB supergravity, this is expected to be the case. It is also interesting to explore the direct consequences of  the twisted conformal symmetry on the corresponding NCOS and in particular features like Hagedorn transition \cite{Gubser:2000mf}. One may also explore extending these considerations about the S-duality and NCOS to the non-unimodular cases and to generalized IIB supergravity.

\section*{Data Management} 
No additional research data beyond the data presented and cited in this work are needed to validate the research findings in this work.

\section*{Acknowledgements}
I.B. is partially supported by the Russian Government program in the 
competitive growth of Kazan Federal University. E.\'O C. is partially supported by the Marie Curie Grant No.\ PIOF-2012- 328625 T-DUALITIES. 
The work of J.S.\ is supported by the Japan Society for the Promotion of Science (JSPS). 
 M.M. Sh-J. is supported in part by the Saramadan Iran  Federation, the junior research chair on black hole physics by the Iranian NSF 
and the ICTP network project NET-68. K.Y.\ acknowledges the Supporting Program for Interaction-based Initiative Team Studies (SPIRITS) 
from Kyoto University and a JSPS Grant-in-Aid for Scientific Research (C) No.\,15K05051. 
This work is also supported in part by the JSPS Japan-Russia Research Cooperative Program 
and the JSPS Japan-Hungary Research Cooperative Program.

\section*{Appendix} 
\setcounter{equation}{0}

\subsection*{Four-dimensional conformal algebra}

We record the conformal algebra $\frak{so}(4, 2)$ employed in this work, 
\bea
\left[ D, P_{\mu} \right] &=& P_{\mu}, \quad \left[ D, K_{\mu} \right] = - K_{\mu}, \nn
\left[ P_{\mu}, K_{\nu} \right]  &=&  2 \left( \eta_{\mu \nu} D +  M_{\mu \nu} \right), \\
\left[ M_{\mu \nu}, P_{\rho} \right] &=& {-} 2 \eta_{\mu [ \nu} P_{\rho]}, \quad  \left[ M_{\mu \nu}, K_{\rho} \right] = {-} 2 \eta_{\mu [ \nu} K_{\rho]}, \nn
\left[ M_{\mu \nu}, M_{\rho \sigma} \right] &=& {-} \eta_{\mu \rho} M_{\nu \sigma} {+} \eta_{\nu \rho} M_{\mu \sigma} {+} \eta_{\mu \sigma} M_{\nu \rho}  {-} \eta_{\nu \sigma} M_{\mu \rho}. \nonumber
\eea
The algebra can be realized in terms of differential operators as 
\bea
P_{\mu} &=& -\partial_{\mu}, \quad K_{\mu} =  -(x_{\nu} x^{\nu}  +z^2) \partial_{\mu} + 2 x_{\mu} (x^{\nu} \partial_{\nu} + z \partial_z), \nn
D &=& - x^{\mu} \partial_{\mu}  - z \partial_{z}, \quad M_{\mu \nu} =  x_{\mu} \partial_{\nu} {-} x_{\nu} \partial_{\mu}. 
\eea

\subsection*{Duality between YB $\sigma$-models and NC sYM for conformal twists} 

In the body of lthis article, we determined $\Theta^{MN}$ for two $r$-matrices $r_1$ and $r_2$. We have conjectured for unimodular $r$-matrices, for example $r_2$, that the YB deformation is dual to a NC deformation of $\mathcal{N} = 4$ sYM. To support this claim, we now show that Eq.\ (30), evaluated at $z=0$, agrees with the resulting NC parameter from the corresponding conformal twist. The analysis presented here generalizes known Drinfeld twists with respect to the Poincar\'e subalgebra to the conformal algebra.

Let us recall the $r$-matrix, 
\be
r_2 = \frac{1}{2} (P_0 - P_3) \wedge (D + M_{03}). 
\ee
We introduce null coordinates, $x^{\pm} = x^0 \pm x^3$, so that the four-dimensional Minkowski metric is 
\be
\dd s^2 = - \dd x^+ \dd x^- + (\dd x^1)^2 + ( \dd x^2)^2. 
\ee
It is worth noting that $\eta_{+ - } = - \frac{1}{2}, \eta^{+-} = -2$. 
In these coordinates, the generators correspond to differential operators: 
\be
P_0 - P_3 = - 2 \partial_{-}, \quad D + M_{03} = - 2 x^+ \partial_{+} - x^1 \partial_1 - x^2 \partial_2. 
\ee
Note, there is no $z$-dependence, and the operators are essentially the AdS$_5$ Killing vectors 
evaluated at $z =0$. Following the standard procedure, we introduce the twist element, which acts on the commutative algebra $\mathcal{A}$ of functions, $f(x), g(x)$, in Minkowski space, 
\be
\mathcal{F} = {\rm e}^{-2 i \eta r_2} =  {\rm e}^{-i \eta (P_0 - P_3) \wedge (D + M_{03})}. 
\ee

The star product then takes the form 
\bea
&& f(x) \star g(x) \nonumber \\ 
&=& m \circ \mathcal{F} (f(x) \otimes g(x))  \nonumber \\
&=& m \circ {\rm e}^{-i \eta (P_0 - P_3) \wedge (D + M_{03})} (f(x) \otimes g(x)) \nn
&=& m \circ {\rm e}^{-i \eta  \partial_{-}  \wedge (2 x^+ \partial_{+} + x^1 \partial_1 + x^2 \partial_2)}( f(x) \otimes g(x)), 
\eea
where $m$ denotes the operation of commutative multiplication, $m (f(x) \otimes g(x) ): = f(x) g(x)$. Taking $f(x) = x^{\mu}, g(x) = x^{\nu}$, $\mu, \nu = +, -, 1, 2$, while expanding to first order, one finds 
\bea
x^{\mu} \star x^{\nu} &=& x^{\mu} x^{\nu}  -  \frac{i}{2} \eta (x^+ \eta^{\mu +} \eta^{\nu -}  - x^1 \eta^{\mu +} \eta^{\nu 1}  \nonumber \\ 
&& \hspace*{2cm} - x^2 \eta^{\mu +} \eta^{\nu 1} - \mu \leftrightarrow \nu), \nn
x^{\nu} \star x^{\mu} &=& x^{\nu} x^{\mu} - \frac{i}{2} \eta (x^+ \eta^{\nu +} \eta^{\mu -}  - x^1 \eta^{\nu +} \eta^{\mu 1} \nonumber \\ 
&& \hspace*{2cm} - x^2 \eta^{\nu +} \eta^{\mu 1} - \nu \leftrightarrow \mu).  
\eea
Therefore, the Moyal bracket is 
\bea
&& [ x^{\mu}, x^{\nu} ]_{\star} =  x^{\mu} \star x^{\nu} - x^{\nu} \star x^{\mu} \nonumber \\
&=&  - i \eta (x^+ \eta^{\mu +} \eta^{\nu -}  - x^1 \eta^{\mu +} \eta^{\nu 1} - x^2 \eta^{\mu +} \eta^{\nu 1} - \mu \leftrightarrow \nu). \nonumber 
\eea
At this stage, it is easy to read off the nonzero components of $\Theta^{\mu \nu}$, 
\be
\Theta^{-+} = - 4  \eta x^+, \quad \Theta^{-1} = - 2  \eta x^1 , \quad \Theta^{-2} = - 2  \eta x^2. 
\ee
This precisely agrees with Eq.\,(30), which was derived from the open string description and evaluated at $z=0$.

\end{document}